# Exploring Regional Development of Digital Humanities Research: A Case Study for Taiwan

Kuang-hua Chen[*] and Bi-Shin Hsueh

Department of Library and Information Science

National Taiwan University

No. 1, Sec. 4, Roosevelt Rd., Taipei 10617, TAIWAN, R.O.C.

khchen@ntu.edu.tw, r01126004@ntu.edu.tw

## Abstract

This study analyzed references and source papers of the Proceedings of 2009-2012 International Conference of Digital Archives and Digital Humanities (DADH), which was held annually in Taiwan. A total of 59 sources and 1,104 references were investigated, based on descriptive analysis and subject analysis of library practices on cataloguing. Preliminary results showed historical materials, events, bureaucracies, and people of Taiwan and China in the Qing Dynasty were the major subjects in the tempo-spatial dimensions. The subject-date figure depicted a long-low head and short-high tail curve, which demonstrated both characteristics of research of humanities and application of technology in digital humanities. The dates of publication of the references spanned over 360 years, which shows a long time span in research materials. A majority of the papers (61.41%) were single-authored, which is in line with the common research practice in the humanities. Books published by general publishers were the major type of references, and this was the same as that of established humanities research. The next step of this study will focus on the comparison of characteristics of both sources and references of international journals with those reported in this article.

Keywords: Descriptive Analysis, Digital Humanities, Subject Analysis, Taiwan

---

[*] Corresponding author

# 1. Introduction

The research topics of digital humanities (DH) have emerged in recent years due to the effort of digital libraries and the interaction of multiple disciplines worldwide. The concept of "digital humanities" has evolved over the years. Digital humanities was first known by the name "humanities computing". McCarty (2003), Svensson (2009), and Kirschenbaum (2010) have discussed "humanities computing," "digital humanities," and the methodologies that may be applied. According to ACH (Association for Computers and the Humanities), a major professional society for the digital humanities, "digital humanities" is a broad term encompassing a wide range of subject domains and communities of practice, including computer-assisted research, pedagogy, and software and content development in humanistic disciplines, such as literature and language studies, history, or philosophy. DH also has been engaged with the relationship between digital technologies and humanities methods and with the ways they may influence each other (ACH, 2013). Another organization, EADH (The European Association for Digital Humanities), has focused on the mission of representing and bringing together the digital humanities in Europe across the entire spectrum of disciplines that apply, develop, and research digital humanities methods and technology (EADH, 2013). In addition, ACH and EADH formed an umbrella organization, ADHO (Alliance of Digital Humanities Organizations), in 2002. Other organizations, CSDH/SCHN (Canadian Society for Digital Humanities/Société canadienne des humanités numériques), centerNet, aaDH (Australasian Association for Digital Humanities), and JADH (Japanese Association for Digital Humanities) joined ADHO in succession. The goals of ADHO are to promote and support digital research and teaching across arts and humanities disciplines, drawing together humanists engaged in digital and computer-assisted research, teaching, creation, dissemination, and beyond, in all areas reflected by its diverse membership (ADHO, 2013). Since then, the development of DH has become much more active globally.

In order to observe the development of DH in Taiwan, one can look back to the joint efforts of projects of digital libraries/museums. The National Science Council in Taiwan initiated the Digital Libraries/Museums Program (DLMP) in 1998, and made considerable effort to digitalize Taiwan's cultural heritage. In 2002, DLMP was followed by a new program entitled National Digital Archives Program (NDAP), whose mission was to preserve Taiwan's cultural heritage in digital form. In fact, the digitization of cultural heritage has become a core task in the country (Hsiang, 2011). In general, more than ten million digital objects have been created since the initiation of DLMP and NDAP. It is now time to consider how to explore the meaning of these digital objects, how to identify underlying structures and trends, how to investigate relationships of digital objects, how to construct contexts of related digital objects, and how to collocate digital objects in tempo-spatial dimensions.

In order to promote research on the aforementioned digital objects and international cooperation, National Taiwan University has held the International Conference of Digital Archives and Digital Humanities (DADH) annually since 2009. The conference offers researchers a venue to share their findings while discussing the progress and future of digital humanities. It is the only conference in Taiwan named with digital humanities and has attracted many professional participants domestically and internationally.

The research issues covered by the DADH conferences have been history, geography, archaeology, sociology, politics, and so on. The applied computing technology has included artificial intelligence, machine learning, digitization, database technology, and internet technology. Although the DADH conference is an international conference, a lot of papers still have dealt with Taiwan or China related issues. Therefore, considering aspects of social sciences, arts and humanities, and computing technology, we could gain some insight on the development of DH in Taiwan by analyzing papers of DADH and their references. By collecting the papers (sources may be used hereafter) of this conference, we analyzed references listed in each paper in regards to the following facets: (1) type of reference, (2) number of authors, (3) date of publication, (4) place of publication, (5) type of publisher, (6) temporal span, (7) spatial coverage, and (8) subject area. The analyses based on the last three facets also were carried out for source papers. Since some data may not be available in sources or references, this will have some impact on this investigation. Nevertheless, the analyses of the source papers, as well as their references, could give us an overall picture of the research and citation characteristics of DH research in Taiwan.

The library practices on cataloguing, descriptive analysis, and subject analysis but not citation analysis, such as co-citation, will be used in this study. This article is structured as follows. Section 2 introduces the DADH conference and its proceedings. Section 3 shows results of descriptive analysis of references of papers of the DADH conference proceedings. Section 4 follows up with the subject analysis of references and sources and discusses the results. Section 5 discusses related works. Section 6 presents the conclusions.

## 2. Briefing DADH Conference

The International Conference of Digital Archive and Digital Humanities (DADH) focuses on issues of processing huge data via digital technology and information analysis methods and on how digital data could facilitate knowledge creation. In addition, this conference would like to promote interaction of humanities research and information technology (DADH, 2011). The diversity of participants from Japan, China, HK, Thailand, UK, USA, *etc*., shows its increasingly international reach. The forum provides an opportunity for discussion among researchers so they can get to know one another and what their colleagues are doing or planning to do. At the same time, they might get some inspiration from one another.

The Digital Archive Research Development Center at National Taiwan University held the first DADH in 2009. It focused on four major issues: histories and databases; data mining and extraction; visualizing narratives; and database production, dissemination, and archiving. Most of the research targets of the first conference dealt with historical archives. With the help of computing technology, quantitative analysis for tremendous records could now be conducted and interpreted in a way not possible with regular human effort.

The second conference, held in 2010, discussed more about core concepts in digital humanities from the basic ideas and technology progress to some database analysis. The research projects emphasized analyses of humanities records with various computing technologies and their practical implications. Due to DH having received considerable attention globally and domestically, the papers also addressed the global usage and some local adoption of various resources for digital humanities.

In the third conference, analyses of archives and documents were the major research issues. Chinese natural language processing, corpus linguistics, and data mining were commonly used to analyze records and archives. In addition, geographical information was another important area that researchers of digital humanities wanted to deal with.

The topics of the fourth conference (2012) followed the similar themes of the past three years. Many technology-related topics were covered in the fourth conference, including data categorization and clustering, visualizing demonstration, text and sentence analysis, time and space, term extraction, and conceptualization.

Table 1 presents general information of the conference proceedings collected by this research project. Each of the proceedings was given a title. Each proceeding was composed of several research topics. Papers with relevant research topics would be put together in a group.

## 3. Descriptive Analysis

There were four proceedings in five volumes published for DADH from 2009 to 2012. Each volume contained about ten or more papers. All of the papers and their respective references were analyzed without considering the nationality of the authors. A total of 1,104 references cited in 59 papers of the proceedings were collected and analyzed. By analyzing these references, we could investigate the citation characteristics of research of digital humanities in Taiwan, even though some authors were not Taiwanese. The first information we could attain from the references concerned the ways contributors made their works, either by authoring, editing, translating, or compiling. Most of the references (93.93%) were authored, and only 49 (4.44%) out of 1,104 were edited. The following focuses on descriptive analysis that was based on descriptive cataloguing of library practices. Five facets have been considered here: type of reference, number of authors, date of publication, place of publication, and type of publisher.

(1) Type of Reference

Table 2 shows the distribution of the types of references. The top three types were book, journal article, and proceedings paper. The "book" was the most popular type, which might possibly be the research tradition in the arts and humanities. That is, the humanities domain tends to cite books because of their informativeness and comprehensiveness.

It was worthy of noting that there were a proportion of references in "Web Page" type. 68 out of 1,104 references were in the format of web page, giving readers a chance to reach the resources online.

(2) Number of authors

Of the 1,104 references, 678 (61.41%) were written by a single author. 70 of the references (6.34%) were written by organizations. This shows that there were several organizations devoting themselves to the study of digital humanities. There were still 19 references (1.72%) left unknown. Table 3 shows the detailed statistics. It seemed that researchers preferred to work alone rather than to cooperate.

(3) Date of Publication

The date of publication of the references ranged from 1654 to 2012. In general, most cited references were published after 2000, but the count of references dropped significantly after the year of 2009. Among 1,104 references, there were 16 references (1.45%) lacking the date of publication. There were two cases where the publication dates were a time span. One was continuously published from 1895-1945, and the other was from 1908 to 1909. Both of these were archival data; thus, one might not be able to assign a single year to them. Figure 1 shows the distribution for the date of publication. The date of publication of cited references presented a curve with both a long-low head and a short-high tail. The long-low head phenomenon was much like the research fields of humanities, where historical materials would be cited, while the short-high tail symbolized the applications of computing technology, where up-to-date research results would be preferred.

(4) Place of Publication

The next statistics deal with the place of publication. Among the 1,104 references, the place of publication for 497 references (45.02%) could not be identified. 411 references (37.23%) were published in Taiwan and China. This was not surprising since most of the papers dealt with China or Taiwan related materials. Table 4 displays the statistics for place of publication of these references. It was interesting that a variety was shown in the place of publication. More than 10 references published in USA, UK, Japan, Australia, Germany, and New Zealand could be found. Through Table 4, one could realize that the place of publication was scattered globally. We may conclude that it demonstrated not only the interdisciplinary nature but also the international citation characteristics of the research area of digital humanities.

(5) Type of Publisher

The last part of the descriptive analysis was about the type of publisher. We focused on general publisher, university, museum/library, and organization. If no information was available, N/A was denoted. Among the 1,104 references, the type of publisher could not be identified 543 times (49.18%). The main type of publisher was general publisher, which published 300 titles (27.17%). University was the second major type of publisher, which published 166 titles (15.04%). The third one was museum/library, which shared a few. Also, there were some organizations around the world publishing materials relevant for digital humanities research. The organizations mentioned here were government agencies, national archive administrations, and academic organizations. Table 5 shows the details.

## 4. Subject Analysis

Section 3 revealed the descriptive characteristics of the references. This could help us attain an understanding of the citation characteristics of digital humanities research. This section will focus on the subject analysis, which was based on subject cataloguing of library practices. Three subject-related facets, temporal span, spatial coverage, and subject area, of the references were identified. Here, subject analysis on source papers also was carried out for comparison.

**4.1 Subject Analysis for References**

(1) Temporal Span

When taking a closer look at the references, the temporal span about which these references were concerned could be identified. The periods of our concern were based on political transitions of central government, since most of the references discussed the modern history of China and Taiwan or nearby places. In fact, not many references pointed out temporal information clearly. Table 6 shows statistics of the temporal span of references, which were recognized either by the titles or by the publishing organizations.

According to Table 6, the time spans of the references could not be identified 906 times (82.07%). 19 (1.72%) of the references were classified as a vague time period due to no clear temporal information being available, which included the past, recent times, and modern times. Most of the references discussed the materials, events, or people of the Qing Dynasty, the Japanese Colonial Period, and the time of the Republic of China (R.O.C.).

(2) Spatial Coverage

Some of the references pointed out the regions covered by their studies. Among 1,104 references, 736 (66.66%) offered spatial information in their titles. As a result, the spatial coverage could be identified easily. According to the data collected from the references, the major locations of concern were Taiwan, China, Pacific regions, and Japan. Taiwan shared 178 counts; China shared 87 counts; the Pacific region shared 48 counts; and Japan shared 25 counts. It was evident that Asia was the focus of research in DADH conferences. Most researchers concerned with Taiwan and China showed their interest in the historical archives, while Pacific regions related research projects mostly were interested in prehistoric Pacific islands culture. Research related to Japan often dealt with GIS (Geographic Information System) issues and some historical and modern Kyoto geographical issues.

(3) Subject Area

In order to investigate the subject areas of DH research, this study categorized the 1,104 references into their related subject areas. Because of the nature of interdisciplinary research for digital humanities, two aspects would be used to identify the subject area of each reference. The first aspect was social sciences and humanities (SSH); the second one was technology (T). Under each major aspect, various minor aspects existed. The minor aspects could be regarded as subject tags, which would be used to tag each reference. In addition, each reference could be assigned more than one subject tag. Table 7 shows subject tags in aspects of SSH and T and shows the tagging results. In total, 1,104 references shared 1,829 tags. As previously mentioned, since more than one tag may be assigned, the total number of tags could be more than the number of references.

Table 7 showed that "history," "politics," "culture," and "literature" were popular subjects in digital humanities. Comparatively, the most popular subject area of the aspect of social sciences was "politics". "Data mining" and "artificial intelligence" were the most often applied technologies when researchers dealt with studies of digital humanities. In addition, each reference had 1.16 SSH-related (social sciences and humanities related) tags, but only 0.50 technology tags. This meant research in digital humanities still focused much more on data or archives than on computing technology. Despite some researchers of social sciences and humanities questioning computing technology, computing technology does help in doing DH research. Nevertheless, computing technology cannot replace the core value of humanities and social sciences' research.

In order to investigate the transition of subject areas over time, the distribution of number of subject tags of references versus their publication date is shown in Figure 2 and Figure 3. Figure 2 provides an overall picture of the distribution, which shows more SSH tags than T tags and shows more tags in recent years. Since most of references were published after 1890, Figure 3 shows a more detailed picture from the year 1890. The cited references with history, politics, and literature tags showed a comparatively long time span in their publication date, which meant history, politics, and literature related issues continuously attracted researchers to explore underlying connections, relationships, contexts among records, archives, and materials.

**4.2 Subject Analysis for Sources**

For the purpose of comparison, 59 papers (sources) presented in the DADH conference were analyzed for subject coverage. The three facets that we used for analysis on references were adopted again.

(1) Temporal Span

Table 8 shows the temporal span of source papers. Among the 59 source papers, only 14

papers (less than a quarter) mentioned the temporal information in their titles. Most of them discussed humanities issues in the Qing Dynasty, and other time periods were distributed variously. Comparing the results for references, the Qing Dynasty was also the major research concern. This may be due to the availability of archives of Qing Dynasty, thus attracting many researchers to work on materials in the Qing Dynasty.

(2) Spatial Coverage

Table 9 shows the spatial coverage of source papers. 21 items (35.59%) of spatial information could be recognized from titles. It was not surprising that Taiwan was the most popular focus, with 11 counts. Japan had 5 counts. It is interesting that one paper was entitled "Creating a Digital Database of Japanese Ceramics in Western Collection". This paper was likely a cross-country research topic. The third one was China with 2 counts. Korea, Afghanistan, and the Pacific region also were topics in source papers of DADH proceedings. With comparison to the results of references, Taiwan still was the most highly studied region, but the second one in reference papers was China, rather than Japan (as in the source papers). This is because considerably more presenters or participants of DADH conference were from Japan than from Mainland China.

(3) Subject Areas

A total of 59 source papers were assigned subject tags in the same way as in the tagging process carried out for reference papers. Table 10 shows the tagging results. "History" was also the core research issue of digital humanities studies in Taiwan. "Data mining" and "digitization" were the top two subject tags in the technology aspect. "General technology" was also popular, which broadly dealt with technology issues in digital humanities. For the social sciences domain, "politics" was still the one with the most counts. Generally speaking, we could find that humanities and technology domains were more popular than social sciences. In addition, reference papers and sources papers showed a few differences in the subject areas they covered. Figure 4 shows the absolute distribution of tags for references (1,829 tags) and sources (156 tags), respectively. Figure 5 shows the relative distribution of tags.

As Figure 5 shows, the relative distribution of tags revealed much more meaningful information than the absolute one. Among the T tags, "digitization" and "database" related technology were more important in sources than in references. Among the SSH tags, "history," "linguistics," and "law" were the three major issues in sources, which was not the case in the references. Nevertheless, it is noted that the number of source papers was insufficient to reach concrete conclusions.

## 5. Related Work

In general, few similar investigations or citation analyses have been carried out for research of digital humanities. Nevertheless, there has been impetus in applying bibliometric methodology to arts and humanities in recent years. In this, citation analysis has been one of the major practices. Knievel and Kellsey (2005) reported results of analyzing 9,131 references of eight humanities fields and found citation patterns varied widely among humanities disciplines. 34.7% of the citations in the art discipline were from foreign language resources, while only 0.3% of the citations were from foreign language resources in the philosophy discipline. The distribution of the foreign language citations also varied between different disciplines and the cited proportion of monographs. The paper mainly examined the use of foreign language resources by the scholars in each field and the relative percentages of books and journals cited. The results showed both were of significant diversity among different humanities disciplines, and the conclusion suggested the uniqueness of every discipline in humanities.

Hellqvist (2010) discussed citation practices in the arts and humanities from a theoretical and conceptual viewpoint. The author concluded that the nature of diversity in humanities disciplines was quite different from other areas. First, it included several sources from the academic as well as the nonacademic world. At the same time, references may not be regarded as an acknowledgement of previous research. Also, the meaning of a citation was highly context-bound; therefore, generalizations about impact or influence were at best tentative. White, Buzydlowski, and Lin (2000) investigated co-cited author Maps for Digital Libraries systems. This research was an early work that applied citation analysis on arts and humanities to constructing a user interface for humanities researchers.

In fact, few studies have focused on citation analysis for research of digital humanities. Leydesforff and Akdag Salah (2010) reported research results of citation analysis on articles of digital humanities. They created a citation map of articles of digital humanities using Web of Science and A&HCI (one database of Web of Science) to investigate the impact of granularity in the scale of the database. Their results showed that digital humanities related articles were cited in a limited domain of journals with focuses on library and information science, computers and literature, and computer application in linguistics.

The study presented in this paper is not similar to the aforementioned works. Actually, no citation analysis, such as bibliographic coupling and co-citation, was applied in this study. In contrast, descriptive analysis and subject analysis of library practices were used. In some sense, the study of uncovering citation patterns in this paper was based on the aforementioned analyses on cited references of DADH papers. In addition, the principle of literary warrant (Hulme, 1911) was used in the tagging process to assure both SSH tags and T tags would be assigned to sources and references. All tags were generated in a data-driven way, which was inspired by Moed (2005) in doing citation analysis. After tagging and analyzing, the research characteristics of digital humanities were revealed via the analyses on both sources and references of DADH papers.

# 6. Conclusions

Digital Humanities is at a stage of quick and broad development. In an era of interdisciplinary research, digital humanities is a fertile research field that has attracted a lot of researchers of various disciplines. This study managed to explore the current patterns of research of digital humanities in Taiwan. 1,104 references of DADH conference papers were analyzed for their descriptive matters and subjects. Subject analysis for 59 source papers of DADH conference proceedings was also carried out for comparison purposes.

The preliminary results showed that a lot of papers were still single-authored. This is because a lot of cited references were humanities-related studies, and as it has been known traditionally humanities-related papers tend to be single-authored. The figure of dates of publication of references versus reference counts spanned over 360 years, which showed historical materials or archives were of importance in research of digital humanities. In addition, this curve with long-low head and short-high tail phenomena showed both characteristics of researches of humanities and applications of technology. Books published by general publishers were the major type of references, and this was the same as that of established humanities' research.

As to subject analysis of reference papers, historical materials, events, bureaucracies, and people of Taiwan and China in the Qing Dynasty were the major subjects of concern in tempo-spatial dimensions. The analysis of source papers showed similar results. Nevertheless, temporal span and spatial coverage of references scattered variously showed broad interest in research of digital humanities. The figure of subject areas over time demonstrated SSH-related subjects scattered broadly and spanned much more in time line, but T-related subjects did not.

Due to this being the first time to explore research and citation characteristics of digital humanities research in Taiwan, there are a few problems that were not solved in this study. For example, some missing data are not patched in this study, *i.e.*, no information about publication date and place or temporal span and spatial coverage. This will, of course, have an impact on the results. Nevertheless, this attempt is still meaningful and insightful, since it uncovered various characteristics of references and sources of DADH conference proceedings. This could be helpful in understanding current states of digital humanities in Taiwan. In order to examine research of digital humanities in a broad view, the next step of this research will focus on the comparisons of characteristics of both sources and references of international journals, *e.g.*, Journal of Digital Humanities, Digital Humanities Quarterly, and Digital Studies/Le champ numérique, with those reported in this article.

## List of abbreviations

aaDH: Australasian Association for Digital Humanities
ACH: Association for Computers and the Humanities
ADHO: Alliance of Digital Humanities Organizations
CSDH/SCHN: Canadian Society for Digital Humanities/Société canadienne des humanités numériques),
DADH: International Conference on Digital Archives and Digital Humanities
DH: Digital Humanities
DLMP: Digital Libraries/Museums Program in Taiwan
EADH: The European Association for Digital Humanities
JADH: Japanese Association for Digital Humanities
NDAP: National Digital Archives Program in Taiwan
SSH: Social Sciences and Humanities
T: Technology

## Competing interests

The author(s) declare that they have no competing interests

## Authors' contributions

Both authors have equal contributions to this work.

## Authors' information

**Kuang-hua Chen** was born in Taipei, Taiwan, R.O.C. He received his B.S. degree in 1986, master degree in 1991, and Ph.D. of Computer Science in 1996, all from National Taiwan University. He joined Department of Library and Information Science at National Taiwan University in 1996. Currently, he is a professor of Library and Information Science and the Associate University Librarian of National Taiwan University. His research interests are Information Retrieval and Evaluation, Citation Analysis and Evaluation, Natural Language Processing, Digital Library, Digital Humanities, and Intelligent Information Systems. He has published more than 100 research papers, 3 book titles, 5 patents, and other publications. He serves as Chief Editor of one journal and joins editorial board of a few journals. He is the member of Library Association of Republic of China (Taiwan), Association for Computational Linguistics and Chinese Language Processing, Chinese Association of Library & Information Science Education, and Institute of Information & Computing Machinery. Detailed information is available at www.lis.ntu.edu.tw/~khchen/.

**Bi-Shin Hsueh** was born in Taipei, Taiwan, R.O.C. She received her B.S. degree in 2012. Currently, she is the master student of Department of Library and Information Science. Her research interests are Digital Library, Digital Humanities, and User Study.

## Acknowledgments

Authors would like to thank Professor Muh-Chyun Tang and Miss Ming-Hsiang Lu for their insightful comments. Authors are grateful to anonymous reviewers for constructive suggestions. This work is partially supported by "The Aim for the Top University Project, Integrated Platform of Digital Humanities" at National Taiwan University in Taiwan.

Illustrations and figures

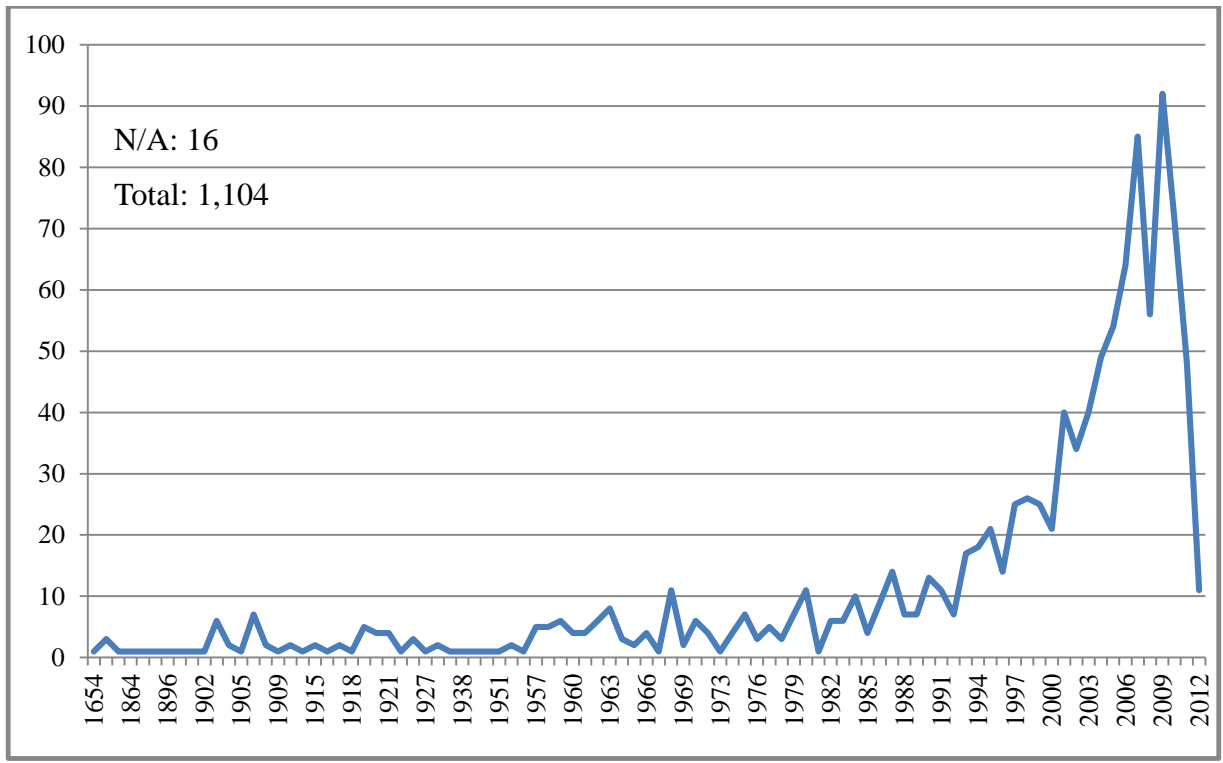

Figure 1. Count of Publication Date of Reference

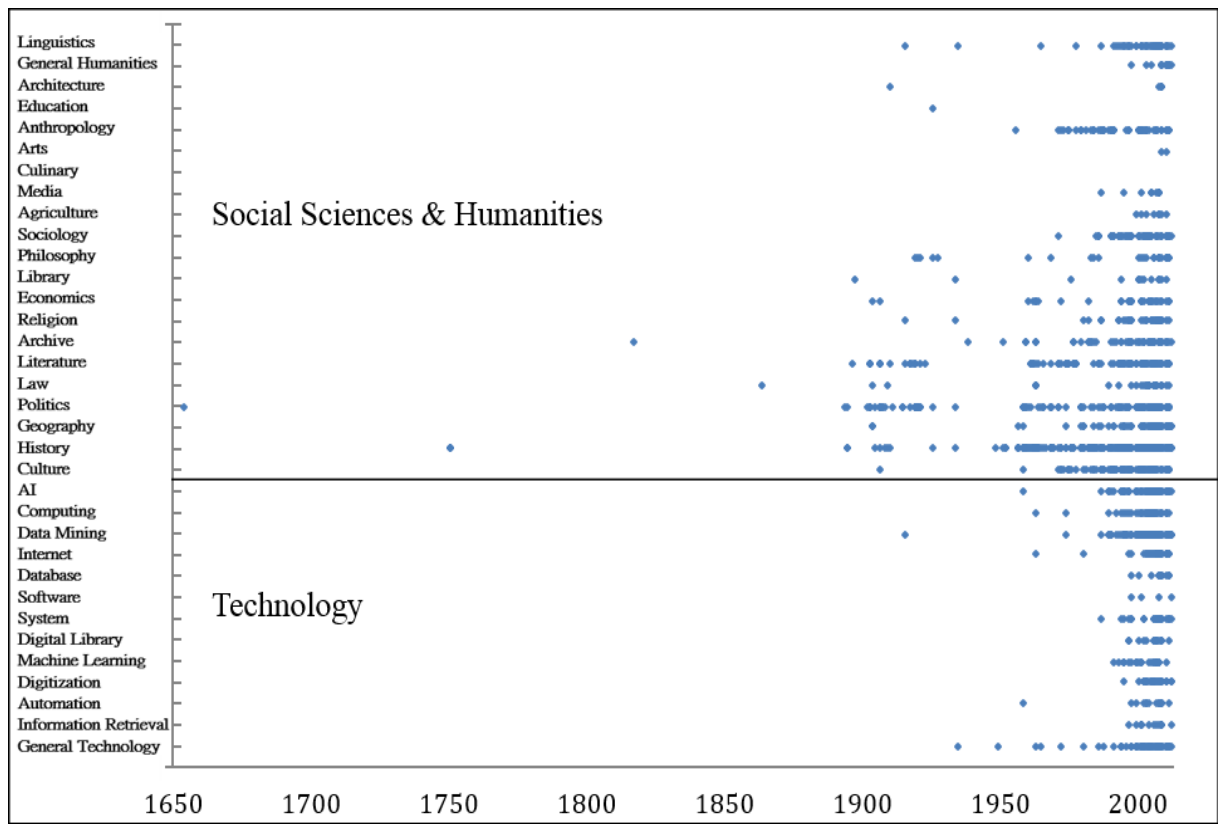

Figure 2. Subject Distribution over Time (1650-2012)

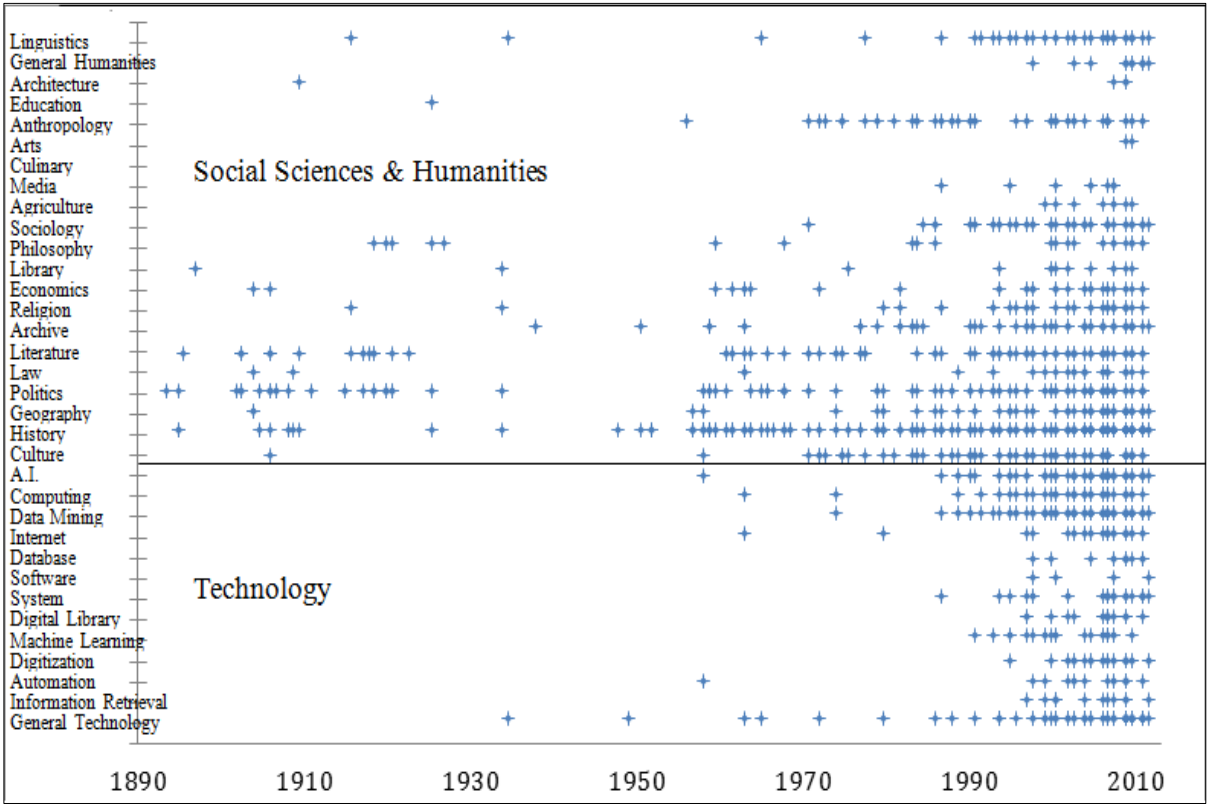

Figure 3. Subject Tags over Time (1890-2012)

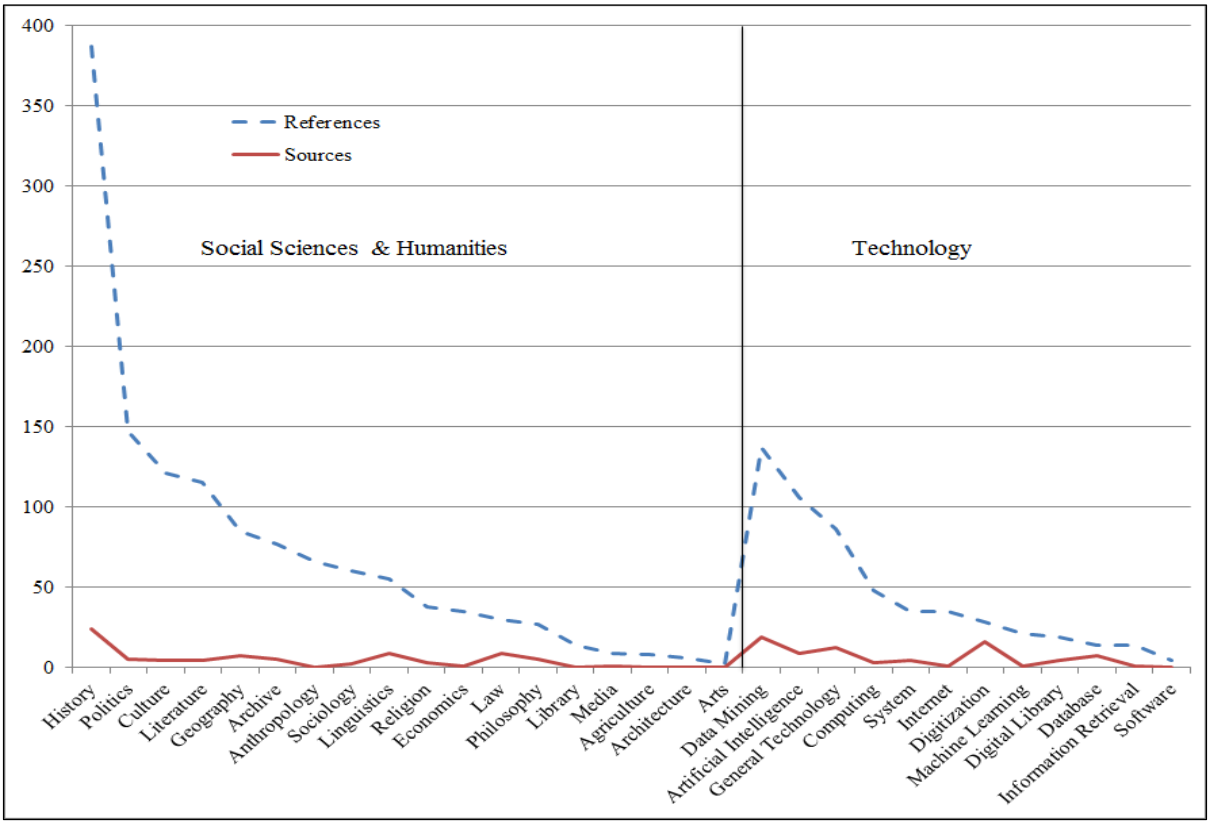

Figure 4. Distribution of Subject Tags of References and Sources

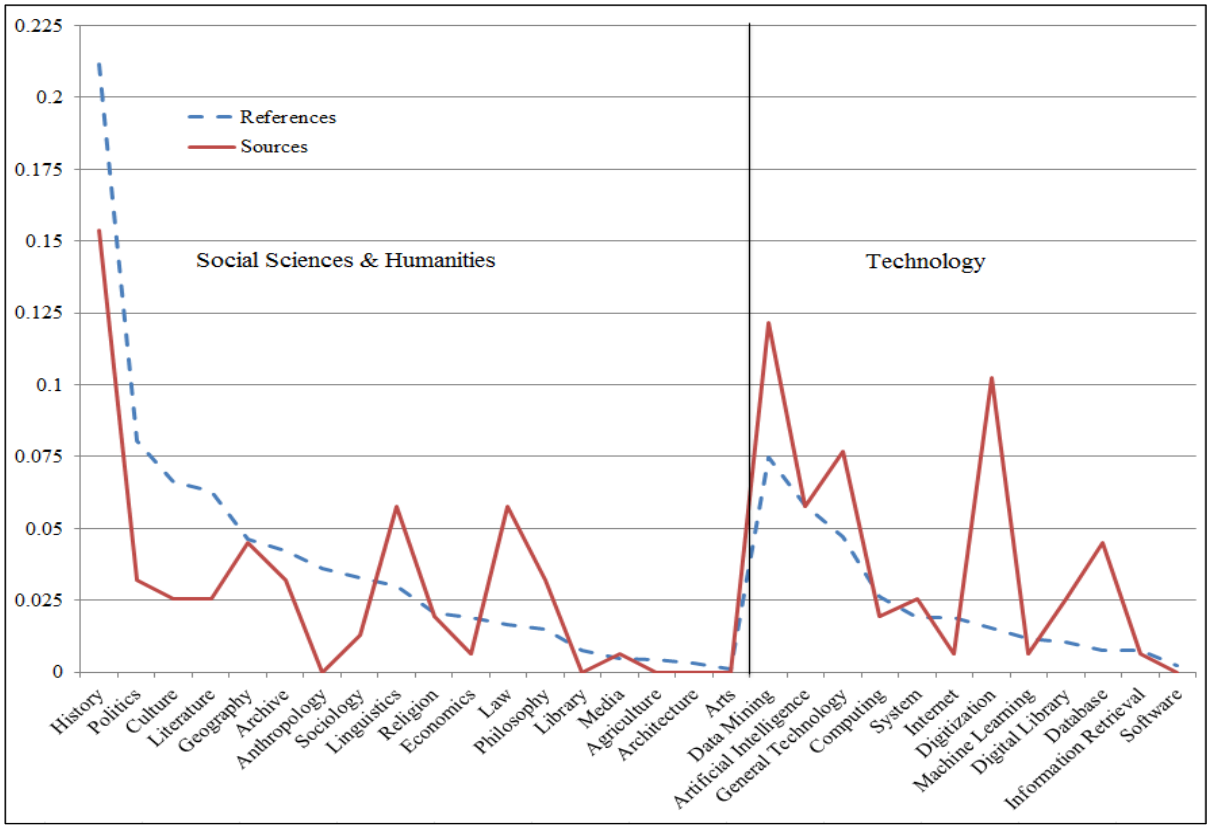

Figure 5. Relative Distribution of Subject Tags of References and Sources

Tables and captions

Table 1. Proceedings of DADH Conference

| Year | 2009 | 2010* | 2010* | 2011 | 2012 |
|---|---|---|---|---|---|
| Title of Proceedings | From Preservation to Knowledge Creation: The Way to Digital Humanities. | New Eyes for Discovery: Foundations and Imaginations of Digital Humanities | Digital Humanities: New Approaches to Historical Studies. | Essential Digital Humanities: Defining Patterns and Paths. | 4th International Conferences of Digital Archives and Digital Humanities |
| Part I | When Historians meet Databases | Back to Basics | Global Approach | Archives & Documents | Classification & Clustering |
| Part II | Data Mining and Extractions | Technologies Forward | Local Adoption | Corpus Linguistics | Visualization |
| Part III | Visualizing Narratives | Ground Truth | Idea Reconsideration | Geographical Information | Literary Analysis |
| Part IV | Production, Dissemination, Archiving | | | | Space & Time |
| Part V | | | | | Term Extraction |
| Part VI | | | | | Conceptual Modeling |

*The proceedings of 2010 conference consists of two volumes.

Table 2. Type of Reference

| Type of Reference | Count | % |
|---|---|---|
| Book (Monograph, Book Chapter, Edited Book) | 479 | 43.39 |
| Journal Article | 309 | 27.99 |
| Proceedings paper | 122 | 11.05 |
| Web Page | 68 | 6.16 |
| Thesis | 48 | 4.35 |
| Magazine | 31 | 2.81 |
| Newsletter/Newspaper | 24 | 2.17 |
| Database | 10 | 0.91 |
| Presentation | 7 | 0.64 |
| Archive | 5 | 0.45 |
| N/A | 1 | 0.09 |
| Total | 1,104 | 100.00 |

Table 3. Number of Authors

| Author | Count | % |
|---|---|---|
| 1 Author | 678 | 61.41 |
| 2 Authors | 171 | 15.49 |
| More than 3 Authors | 162 | 14.68 |
| Organization | 70 | 6.34 |
| Unknown | 19 | 1.72 |
| Project | 3 | 0.27 |
| N/A | 1 | 0.09 |
| Total | 1,104 | 100.00 |

Table 4. Place of Publication

| Place of Publication | Count | Place of Publication | Count | Place of Publication | Count |
|---|---|---|---|---|---|
| Taiwan | 294 | Australia | 17 | Papua New Guinea | 1 |
| China | 117 | Germany | 15 | Canada | 1 |
| USA | 80 | New Zealand | 14 | Italy | 1 |
| UK | 36 | France* | 7 | Switzerland | 1 |
| Japan | 20 | Czech Republic | 3 | N/A | 497 |
| Total | | | 1,104 | | |

*3 references published in New Caledonia were added to France.

Table 5. Type of Publisher

| Publisher | Count | % |
|---|---|---|
| General Publisher | 300 | 27.17 |
| University | 166 | 15.04 |
| Museum/Library | 5 | 0.45 |
| Organization* | 90 | 8.15 |
| N/A | 543 | 49.19 |
| Total | 1,104 | 100.00 |

*Government agencies and institutions were regarded as "Organization".

Table 6. Temporal Span of References

| Period | Count | Period | Count | Period | Count |
|---|---|---|---|---|---|
| Prehistory | 1 | Ming Dynasty | 1 | Late Qing Dynasty, Beginning of Republic of China | 10 |
| Han Dynasty | 3 | Late Ming Dynasty, Early Qing Dynasty | 1 | Republic of China | 14 |
| Eastern Han Dynasty | 2 | Ming & Qing Dynasty | 17 | Japanese Colonial Period | 21 |
| Wei, Jin, and the Northern and Southern Dynasties | 1 | Qing Dynasty | 100 | Vague Time Period* | 19 |
| Song Dynasty | 1 | Late Qing Dynasty | 7 | N/A | 906 |
| Total | | | | | 1,104 |

*Vague time period were ancient times, recent ages, or modern ages.

Table 7. Subject Areas of References

| Subject Area | Tag | Number of tags | % | Rank |
|---|---|---|---|---|
| Social Sciences & Humanities | History | 387 | 35.05 | 1 |
| | Politics | 147 | 13.32 | 2 |
| | Culture | 121 | 10.96 | 4 |
| | Literature | 115 | 10.42 | 5 |
| | Geography | 85 | 7.70 | 8 |
| | Archive | 77 | 6.97 | 9 |
| | Anthropology | 66 | 5.98 | 10 |
| | Sociology | 60 | 5.43 | 11 |
| | Linguistics | 55 | 4.98 | 12 |
| | Religion | 38 | 3.44 | 14 |
| | Economics | 35 | 3.17 | 15 |
| | Law | 30 | 2.72 | 18 |
| | Philosophy | 27 | 2.45 | 20 |
| | Library | 14 | 1.27 | 23 |
| | Media | 9 | 0.82 | 26 |
| | Agriculture | 8 | 0.72 | 27 |
| | Architecture | 6 | 0.54 | 28 |
| | Arts | 2 | 0.18 | 30 |
| Technology | Data Mining | 137 | 12.41 | 3 |
| | Artificial Intelligence | 106 | 9.60 | 6 |
| | General Technology | 86 | 7.79 | 7 |
| | Computing | 48 | 4.35 | 13 |
| | System | 35 | 3.17 | 15 |
| | Internet | 35 | 3.17 | 15 |
| | Digitization | 28 | 2.54 | 19 |
| | Machine Learning | 21 | 1.90 | 21 |
| | Digital Library | 19 | 1.72 | 22 |
| | Database | 14 | 1.27 | 23 |
| | Information Retrieval | 14 | 1.27 | 23 |
| | Software | 4 | 0.36 | 29 |
| Total | | 1,829 | 165.67* | |

*One paper could be assigned more than one tag.

Table 8. Temporal Span of Sources

| Temporal Span | Count | Temporal Span | Count | Temporal Span | Count |
|---|---|---|---|---|---|
| Qing Dynasty | 9 | Ming & Qing Dynasty | 1 | Japanese Colonial Period | 1 |
| Prehistory | 1 | Ming & Qing Dynasty & Japanese Colonial Period | 1 | Vague Time Period* | 1 |
| N/A | 45 | Total | | 59 | |

*Vague time period here was modern ages.

Table 9. Spatial Coverage of Sources

| Spatial Coverage | Count | Spatial Coverage | Count | Spatial Coverage | Count |
|---|---|---|---|---|---|
| Taiwan | 11 | China | 2 | Korea | 1 |
| Japan* | 5 | Afghanistan | 1 | Pacific region | 1 |
| N/A | 38 | Total | | 59 | |

*One of the counts in Japan is not only about Japan but also about Western civilization.

Table 10. Subject Areas of Sources

| Subject Area | Tag | Number of tags | % | Rank |
|---|---|---|---|---|
| Social Sciences & Humanities | History | 24 | 40.68 | 1 |
| | Linguistics | 9 | 15.25 | 5 |
| | Law | 9 | 15.25 | 5 |
| | Geography | 7 | 11.86 | 8 |
| | Archive | 5 | 8.47 | 10 |
| | Philosophy | 5 | 8.47 | 10 |
| | Politics | 5 | 8.47 | 10 |
| | Culture | 4 | 6.78 | 13 |
| | Literature | 4 | 6.78 | 13 |
| | Religion | 3 | 5.08 | 17 |
| | Sociology | 2 | 3.39 | 19 |
| | Media | 1 | 1.69 | 20 |
| | Economics | 1 | 1.69 | 20 |
| Technology | Data Mining | 19 | 32.20 | 2 |
| | Digitization | 16 | 27.12 | 3 |
| | General Technology | 12 | 20.34 | 4 |
| | Artificial Intelligence | 9 | 15.25 | 5 |
| | Database | 7 | 11.86 | 8 |
| | System | 4 | 6.78 | 13 |
| | Digital Library | 4 | 6.78 | 13 |
| | Computing | 3 | 5.08 | 17 |
| | Internet | 1 | 1.69 | 20 |
| | Machine Learning | 1 | 1.69 | 20 |
| | Information Retrieval | 1 | 1.69 | 20 |
| Total | | 156 | 264.41* | |

*One source paper could be assigned more than one tag.